\documentclass[journal=jacsat,manuscript=article]{achemso}
\usepackage{times,amsmath}
\usepackage{epsfig}
\usepackage{color}
\usepackage{longtable}
\usepackage{graphicx}% Include figure files
\usepackage{dcolumn}% Align table columns on decimal point
\usepackage{bm}% bold math
\usepackage{tabularx}% bold math
\usepackage{hyperref}
\usepackage{multirow}
\hypersetup{citecolor=black, filecolor= black, linkcolor= black, urlcolor= black}

\title{High Dielectric Ternary Oxides from Crystal Structure Prediction and High-throughput Screening}

\author{Jingyu Qu}
\affiliation{College of Science, China Agricultural University, Beijing, 100083, China}
\alsoaffiliation{Department of Physics and Astronomy, University of Nevada, Las Vegas, NV, 89154, USA}

\author{David Zageceta}
\affiliation{Department of Physics and Astronomy, University of Nevada, Las Vegas, NV, 89154, USA}

\author{Weiwei Zhang}
\email{06010@cau.edu.cn}
\affiliation{College of Science, China Agricultural University, Beijing, 100083, China}

\author{Qiang Zhu}
\email{qiang.zhu@unlv.edu}
\affiliation{Department of Physics and Astronomy, University of Nevada, Las Vegas, NV, 89154, USA}

\begin{document}
%\preprint{APS/123-QED}
\begin{abstract}
The development of new high dielectric materials is essential for advancement in modern electronics. Oxides are generally regarded as the most promising class of high dielectric materials for industrial applications as they possess both high dielectric constants and large band gaps. Most previous researches on high dielectrics were limited to already known materials. In this study, we conducted an extensive search for high dielectrics over a set of ternary oxides by combining crystal structure prediction and density functional perturbation theory calculations. From this search, we adopted multiple stage screening to identify 440 new low-energy high dielectric materials. Among these materials, 33 were identified as potential high dielectrics favorable for modern device applications. Our research has opened an avenue to explore novel high dielectric materials by combining crystal structure prediction and high throughput screening. 
\end{abstract}

\section*{Background \& Summary}

A dielectric is an insulator that becomes polarized under the influence of an applied electric field. Materials exhibiting a high dielectric constant ($\epsilon$) are promising for energy storage applications. 
As an example, a parallel-plate capacitor's energy storage capability is approximately expressed as $U=\frac{1}{2}\epsilon\epsilon_0 E^2$, where $\epsilon_0$ and $\varepsilon$ are the permittivity of vacuum and the material dielectric constant respectively, and $E$ is the applied electric field. 
Materials with high dielectric constants, compared to materials with lower dielectrics, have the potential to store more charge per unit volume, which is critical to high-performance device fabrication as well as miniaturization. Many high dielectric materials, such as ZrO$_2$ \cite{panda2013growth}, HfO$_2$ \cite{nahar2007study}, Al$_2$O$_3$ \cite{jeon2002electrical,gerritsen2005evolution}, Y$_2$O$_3$ \cite{kwo2001properties}, SrTiO$_3$ \cite{tsurumi2002dielectric} and BaTiO$_3$ \cite{moriwake2010first}, have been extensively applied in microelectronic technologies. However, currently used dielectrics are inhibiting the development of cheaper and more efficient devices. 
For example, BaTiO$_3$ applied to multi-layer ceramic capacitors (MLCC) encounters certain limitations concerning the continuing miniaturization of circuit components, which require thinner dielectric layers while retaining the reliability of current advanced capacitors. 
Decreasing the particle size of BaTiO$_3$ introduces a so-called ``size effect'' wherein the ferroelectricity reduces with decreasing particle size and then vanishes below a specific critical size \cite{wang2003grain}. 
There also exists a technical challenge when implementing SiO$_2$ in complementary metal–oxide–semiconductor (CMOS) and dynamic random-access memory (DRAM) devices. When the thickness of SiO$_2$ is reduced to a few nanometers, leakage current increases greatly because of the quantum tunneling \cite{lee2008first}. The bandgap ($E_g$) is also a key property affecting device performance. 
In flash memory, a large band gap is necessary to satisfy the stringent leakage current specification (0 $\sim$ 10$^{-9}$ A $\cdot$ cm$^{-1}$) \cite{yim2015novel}. 
However, there exists a general inverse correlation between $\epsilon$ and $E_g$. Thus, the exploration of new high dielectrics requires a careful balance between the bandgap and dielectric constant.

To date, there are only a few hundred known materials with measured dielectric constants; this includes both organic and inorganic materials. The dielectric constants of the vast majority known inorganic compounds $\left(\sim30,000\right)$ are currently unknown. Modern quantum mechanical calculations provide complementary approaches to experiments with far lower costs. In exploring new materials with high dielectric constants, high-throughput screenings of candidate materials based on density functional theory (DFT) calculations have become popular recently. Utilizing a high-throughput setting, Yim \textit{et al}. \cite{yim2015novel} calculated properties for more than 1800 structures of binary and ternary oxides from the Inorganic Crystal Structure Database (ICSD) and generated a total property map of band gap versus dielectric constant. Petousis \textit{et al}. \cite{petousis2017high,petousis2016benchmarking} developed a computational infrastructure to perform high-throughput screening of dielectric materials based on Density Functional Perturbation Theory (DFPT) and constructed a database of dielectric tensors consisting of 1,056 inorganic ordered compounds. However, these studies focused on only already known materials. Recent successes in crystal structure prediction (CSP) has shown that it is possible to predict new materials prior to synthesis \cite{oganov2019structure}. 
Sharma et al. \cite{sharma2014rational} designed organic polymer dielectrics using a strategy of hierarchical modeling, and their efforts led to the successful synthesis of several new high-$\epsilon$ polymers. Zeng \textit{et al}. \cite{zeng2014evolutionary,zhang2014high} applied CSP methodology to explore high dielectrics in particular systems: hafnia-based oxides and Zr$_x$Si$_{1-x}$O. These results inspired us to explore potential high dielectrics in an expanded chemical space.

In this study, we chose to focus on finding new ternary oxides possessing both high $\epsilon$ and $E_g$, given that binary oxides have already been well studied \cite{kim2016machine,van2018high}. In this work, we combined high throughput calculations with crystal structure prediction methods to screen for target materials in a broad chemical space. Specifically, we chose chemical systems based on the combination of two types of metal oxides between group IA/IIA/IIA and group IVB, namely Ca(Be, Mg, Sr, Ba)O-Ti(Hf, Zr)O$_2$, Al(Ga, In)$_2$O$_3$-Ti(Hf, Zr)O$_2$, and Si(Ge)O$_2$-Ti(Hf, Zr)O$_2$). For each system of A$_m$O$_n$-BO$_2$ (A: IIA/IIIA/IVA; B: IVB), we performed variable composition CSP calculations to search for low energy structures which are likely to be (meta)stable if they can be synthesized. The low energy structures were then extracted and fed to our newly developed computational pipeline to screen their dielectric and band gap properties. As a result, we have provided a list of hypothetical materials which are favorable for high dielectric applications.

\section*{Methods}
\subsection*{Theory and definitions}

There are two mechanisms that contribute to a materials dielectric tensor: the ionic contribution which is a consequence of atomic displacement, and the electronic contribution which is a consequence of electron cloud distortion. As a result, the dielectric tensor can be represented as a sum of these two mechanisms $\epsilon_{\alpha\beta} =  \epsilon^0_{\alpha\beta} + \epsilon^{\infty}_{\alpha\beta} $, where $\alpha$ and $\beta$ denote the directions of the applied electric field and the resulting polarization in the Cartesian coordinate system. In most cases, the electronic contribution $\epsilon^{\infty}_{\alpha\beta}$ is much smaller than the ionic part and can be somewhat disregarded. While, the ionic dielectric tensor component $\epsilon^0_{\alpha\beta}$, due to the atomic displacements in the crystalline unit cell, is much more pronounced. Following Ref. \cite{gonze1997dynamical}, we can obtain the $\epsilon^0_{\alpha\beta}$ value by summing the contribution from each vibrational phonon mode, 
\begin{equation}
\label{eps}
\epsilon^0_{\alpha\beta} = \frac{4\pi}{\Omega_0}\sum_{m} \frac{S_{m,\alpha\beta}}{\omega^2_m}, 
\end{equation}

where $\Omega_0$ is the volume of the primitive cell, $\omega_m$ denotes the frequency of each vibration mode. ${S_{m,\alpha\beta}}$ is the mode-oscillator strength tensor and can be obtained by 
 
\begin{equation}
\label{osci}
S_{m, \alpha \beta}=\left(\sum_{\kappa \alpha^{\prime}} Z_{\kappa, \alpha \alpha^{\prime}}^{*} U_{m}^{*}\left(\kappa \alpha^{\prime}\right)\right)\left(\sum_{\kappa^{\prime} \beta^{\prime}} Z_{\kappa^{\prime}, \beta \beta^{\prime}}^{*} U_{m}\left(\kappa^{\prime} \beta^{\prime}\right)\right),
\end{equation}

where $Z_{\kappa, \alpha \beta}^{*}$ is the Born effective charge and $U_{m}(\kappa\beta)$ are eigen-displacements. The eigen-displacements form an orthonormal basis as shown in the following equation. 
\begin{equation}
\sum_{\kappa \beta} M_{\kappa}\left[U_{m}(\kappa \beta)\right]^{*} U_{n}(\kappa \beta)=\delta_{m n}, 
\end{equation}

where $M_\kappa$ is the mass of the ion $\kappa$. It is obvious $\epsilon$ is dependent on various fundamental quantities which include: Born effective charge, ionic mass, phonon frequencies, and cell volume. A detailed analysis of the influence of these quantities on both $E_g$ and $\epsilon$ is provided in Data records section.

When dielectrics are applied to electronic devices, both dielectric constant and bandgap need to be considered. For our screening, we used a fitness model proposed in a recent work \cite{zeng2014evolutionary}:
\begin{equation}
\label{fitness}
\textrm{max}F_{ED}= 8.1882 ~\textrm{J cm}^{-3} \times \epsilon (E_g/E_{gc})^{2\alpha},    
\end{equation}

where $\alpha$ is 1 for an insulator and 3 for semiconductors, E$_g$ is the bandgap and E$_{gc}$ is 4 eV, which represents the critical value to distinguish semiconductors and insulators. %More description about the equation can be found in the earlier paper\cite{zeng2014evolutionary}. 
Using this descriptor, one can estimate the level of energy storage for any given material.

\subsection*{Workflow}
\begin{figure*}[ht]
\centering
\includegraphics[width=0.9\linewidth]{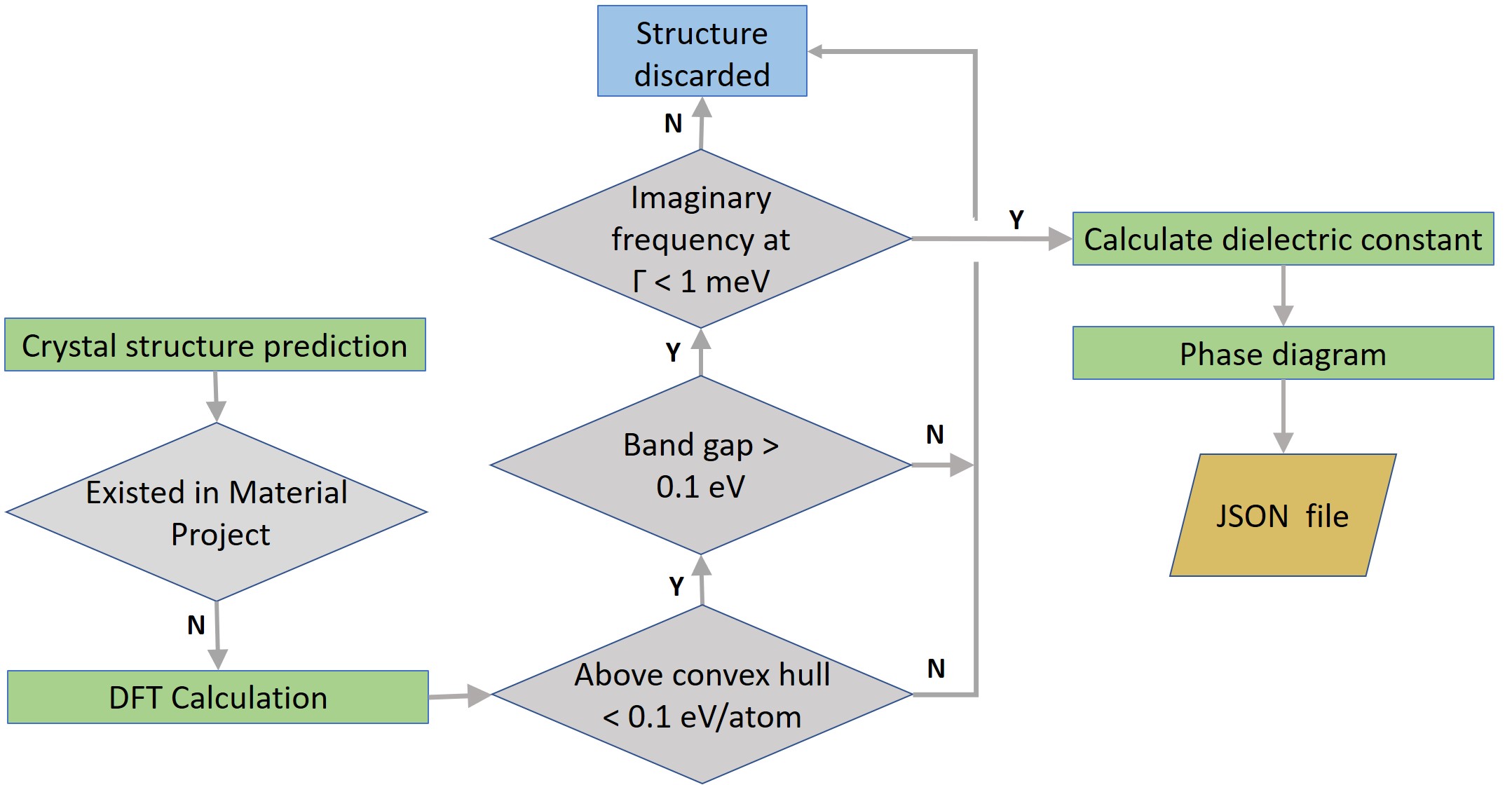}
\caption{\textbf{Flowchart summarizing the calculation method and process.}}
\label{fig:figure1}
\end{figure*}

We performed the screening work by following the flowchart as summarized in Figure \ref{fig:figure1}. The first step is to generate new structures of A$_x$O$_{1-x}$-BO$_2$ from structure prediction. The obtained structures were then compared with the Materials Project database \cite{jain2013}. Since the dielectric properties of structures available in the Materials Project database have been screened thoroughly, we focused on the new structures generated from CSP. The formation energy, band gap, and dielectric constants of the new structures were further calculated by DFT with higher accuracy. To avoid massive calculations, we considered only the structures satisfying the following conditions: (1) within 0.1 eV/atom from the convex hull, (2) bandgap $>$ 0.1 eV. We then calculated the phonon spectrum based on DFPT. Those materials with imaginary frequency at $\Gamma$ point (greater than 1 meV) were also discarded. %Phase diagrams were drawn to represent the possible synthesis pathway in experiment and coexistence of different compounds. Finally, the correlation between bandgap and dielectric constant was analyzed thoroughly.

\subsubsection*{Crystal Structure Prediction }
We utilized the USPEX code \cite{glass2006uspex} to search for new materials within the chemical space described above. For each system (e.g., CaO-TiO$_2$), we performed a CSP calculation with 50 individuals in each generation for 18 generations in total. We initiate the first generation of structures with structures of random composition and space group. The structures in the subsequent generations were generated according to the following variation operators heredity (30\%), random (30\%), mutation (20\%) and permutation (20\%), respectively. The maximum number of atoms in the unit cell is constrained to 24.  

\subsubsection*{Structure Optimization in DFT}
For each structure generated from CSP, DFT calculation was performed with the Vienna ab intio software package (VASP) \cite{kresse1996software}, using the all-electron projector wave (PAW) method \cite{blochl1994projector,kresse1999ultrasoft}. The exchange-correlation energy is treated with the generalized gradient approximation (GGA) within the Perdew-Burke-Ernzerhof (PBE) framework \cite{perdew1996generalized}. Preparation of input parameters and data extraction were done by the Python Materials Genomics (Pymatgen) package \cite{ong2013python}. The structures were fully relaxed until %the atomic force is lower than 0.01 eV\AA$^{-1}$ and
the stress tensor is less than 3 kbar. %Relaxation is important because the dielectric polarization is caused by change of electronic configuration due to atomic vibrations. Incomplete relaxation may result in a large imaginary frequency of acoustic mode. 
The fitness function defined in our structure prediction is the negative of the ab initio free energy of the locally optimized structure. It was found that observed metastable phases are usually not more than 0.1-0.2 eV/atom higher in energy than the ground-state structure \cite{sun2016thermodynamic}. Metastable structures may exist at ambient conditions by means of specific techniques such as doping \cite{lee2008first} or in nano crystalline states \cite{banfield1998thermodynamic}. Herein structures with fitness values higher than 0.1 eV were discarded. All of the obtained structures were fully relaxed until the interatomic forces and the total energies (per atom) are smaller than 0.01 eV/\AA\ and 10$^{-6}$ eV respectively. The plane-wave basis sets have a kinetic energy cutoff of 600 eV. The density of k-mesh by reciprocal volume was set to be 200 in Pymatgen. 

\subsubsection*{Dielectric Constant and Bandgap Calculation}
The DFPT methodology, as implemented in the VASP code \cite{PhysRevB.73.045112} was applied to predict dielectric constants. This method has been widely used to calculate dielectric constants \cite{petousis2016benchmarking,petousis2017high} and refractive indices \cite{naccarato2018designing}. The dielectric tensor is composed of two parts: electronic and ionic contributions. Since the dielectric response calculated by DFPT corresponds only a monocrystalline material, for simplicity, we adopted the same approximation method proposed in Petousis' work \cite{petousis2017high}, in which the polycrystalline dielectric constant was estimated by averaging the eigenvalues of a monocrystalline dielectric tensor. A large bandgap is another important indicator when selecting industrial high dielectrics. The reported material bandgaps were calculated through DFT using the generalized gradient approximation (GGA). It is worth mentioning that a K-point mesh of high density is required for calculation of $\epsilon$, so the K-point reciprocal density is set to be 300 in Pymatgen, which is sufficient to reach the required computation accuracy.

\subsection*{Code availability}

All reported crystal structure prediction calculations were performed using the USPEX code, which is based on evolutionary algorithms to predict structures with only elemental information \cite{glass2006uspex,lyakhov2013new}. Relaxation of structures and DFPT method were carried out by the VASP code \cite{kresse1996software}.

\section*{Data Records}

\subsection*{File format}
The input files containing important parameters and calculation results are stored in a JSON file, which has been uploaded \cite{figshare}. For each material, one can check the properties by accessing values through keys, such as ``e\underline{\hbox to 2mm{}}poly", “e\underline{\hbox to 2mm{}}total" and ”e \underline{\hbox to 2mm{}}electronic", corresponding to polycrystalline dielectric, total dielectric tensor and electronic contribution tensor, respectively (referred to Table \ref{table1}). Other parameters can be found by accessing the “meta” key as listed in Table \ref{table2}.

\begin{table}[ht]
\centering
	\caption{Description of data keys in JSON file}
\begin{tabular}{|l|l|l|}
\hline
 Key & Datatype &  Description  \\ \hline
 e\underline{\hbox to 2mm{}}total    &  array  &  total dielectric tensor \\ \hline          
 e\underline{\hbox to 2mm{}}electronic  &  array &  electronic contribution of dielectric tensor    \\ \hline          
 e\underline{\hbox to 2mm{}}poly    &  numeric    &  dielectric constant of polycrystalline    \\ \hline  
 bornchrg & list & Born effective charge of each ion \\ \hline
 mode & list & frequency, IR intensity and dielectric constant of each mode  \\ \hline
 eigenvalue & list &  eigenvalues of the dynamical matrix \\ \hline
 eigenvector & list & eigenvectors of the dynamical matrix \\ \hline
\end{tabular}
\label{table1}
\end{table}

\begin{table}[ht]
\centering
	\caption{Description of metadata keys in JSON file}
\begin{tabular}{|l|l|l|}
\hline
Key                & Datatype & Description                         \\ \hline
space group number & numeric   & space group number of structure     \\ \hline
point group        & numeric   & point group of structure            \\ \hline
bandgap            & numeric   & bandgap (E$_g$) of material         \\ \hline
energy             & numeric   & DFT calculation energy              \\ \hline
K\_density       & numeric & density of K points in dielectric calculation \\ \hline
POSCAR             & string   & crystal structure in VASP format    \\ \hline
formula            & string   & chemical formula and multiplicative factor                    \\ \hline
POTCAR             & string   & pseudopotential used in VASP        \\ \hline
INCAR              & string   & parameter of dielectric calculation \\ \hline
K\_path            & string   & kpoint path in first brillouin zone \\ \hline
MP\underline{\hbox to 2mm{}}ID & string & Materials project ID of compound \\ \hline     
\end{tabular}
\label{table2}
\end{table}

\subsection*{Effect of cations on E$_g$ and $\epsilon$}
Though there exist many factors which may affect a material's dielectric properties, we have inferred a general trend by analyzing the data from our simulation. In finding this trend, we chose a few groups of structures belonging to the same prototype with different chemical substitutions of metal cations. In a ternary oxide A$_m$O$_n$-BO$_2$ (A: IA/IIA/IIIA; B: IVB) system, we considered the roles of A and B sites separately.

\begin{figure*}[ht]
\centering
\includegraphics[width=0.9\linewidth]{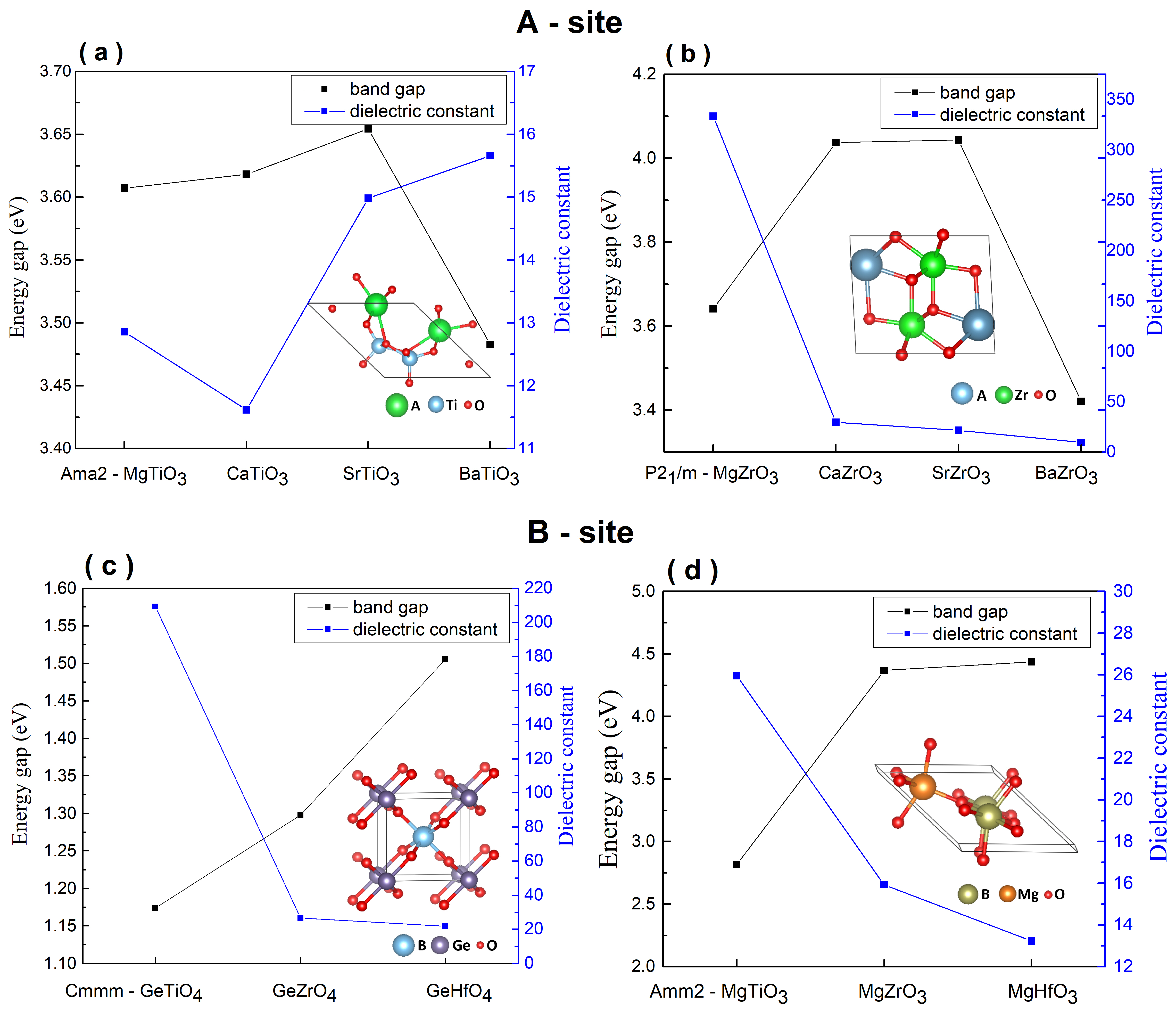}
\caption{\textbf{Dielectric constants and bandgap plots of different sets of compounds.} (a) $Ama2$ IIA-TiO$_3$ (b) $P2_1$/$m$ (c) $Cmmm$ Ge$B$O$_4$ and (d) $Amm2$ Mg$B$O$_3$ structures ($B$=Ti, Zr, Hf). IIA-ZrO$_3$ (IIA=Mg, Ca, Sr, Ba).}
\label{fig:figure2}
\end{figure*}

For the A site, we chose two groups of materials, $Ama2$ IIA-TiO$_3$ and $P2_1$/$m$ IIA-ZrO$_3$. As seen in Figure \ref{fig:figure2}a, b. there is not any obvious correlation between the band gap and the dielectric constant for these two groups of materials. In Figure \ref{fig:figure2}a, IIA-TiO$_3$ ($A$ = Mg, Ca, Sr, Ba) the reported structures are orthorhombic. The bandgap increases as the cation atomic number increases from Mg to Sr and then drops down for the Ba cation. A similar trend of bandgap can also be found in Figure \ref{fig:figure2}b. The dielectric constants in both groups follow completely different trends. This indicates that the factors affecting dielectric tensor are complicated. A more in-depth discussion of this complexity will follow.

For the B site, the dielectric constant and band gap of Ti/Zr/Hf-based oxides share a similar trend regardless of the difference of the cation. In Figure \ref{fig:figure2}c, three materials - GeTiO$_4$, GeZrO$_4$ and GeHfO$_4$ share the same prototypical structure in orthorhombic $Cmmm$ symmetry. Their bandgaps increase as the atomic radius increases; however, the dielectric constants follow an inverse trend. This phenomenon was also verified by the comparison of three orthorhombic $Amm2$ MgBO$_3$ (B = Ti, Zr, Hf) structures in Figure \ref{fig:figure2}d. It is worth noting that dielectric constant of GeTiO$_4$ (209.11) differ significantly from GeZrO$_4$ (26.63) and GeHfO$_4$(21.80) in spite of their structural similarity, demonstrating that the composition of materials can easily influence polarization.

To better understand the atomic mechanism, we analyzed the primary contribution due to ionic vibration for all three materials in GeBO$_4$. From Eq. \ref{osci}, we know the ionic dielectric contribution is not only related to mode frequency but also the Born effective charges ($Z^*$), which are listed in table \ref{table3} for GeBO$_4$ compounds. The $Z^*$ values for Ge in all compounds are larger than their nominal charges (+4 a.u.), indicating the mixed covalent and ionic character of Ge atoms \cite{xie2012electronic}. The decrease of $Z^*$ values of Ge as the increase of the atomic number of B indicates the covalent character of Ge atoms is in order of GeTiO$_4 >$ GeZrO$_4 >$ GeHfO$_4$. As for the B-site, the extraordinarily large value of $Z^*$ indicates the strong covalent bond with oxygen atoms. For GeZrO$_4$ and GeHfO$_4$, the values of $Z^*$ for Zr and Hf are very similar while Ti atoms have notably larger $Z^*$ values. In essence, there is a general trend that $Z^*$(Ti) $>$ $Z^*$(Zr) $>$ $Z^*$(Hf). The largest value of Z* appears at Ti atom in GeTiO$_4$. Xie \cite{xie2012electronic} et al. studied Born effective charges in perovskite Ba$M$O$_3$ ($M$= Ti, Zr, Hf, Sn) and reported the same variation trend on both bandgap and Born effective charge. The variation of Born effective charge can be attributed to the dynamic charge transfer between the transition metal $d$ orbitals and oxygen 2$p$ orbitals by analyzing the deformation of Maximally-Localized Wannier Functions \cite{xie2012electronic} . 

\begin{figure*}[ht]
\centering
\includegraphics[width=0.9\linewidth]{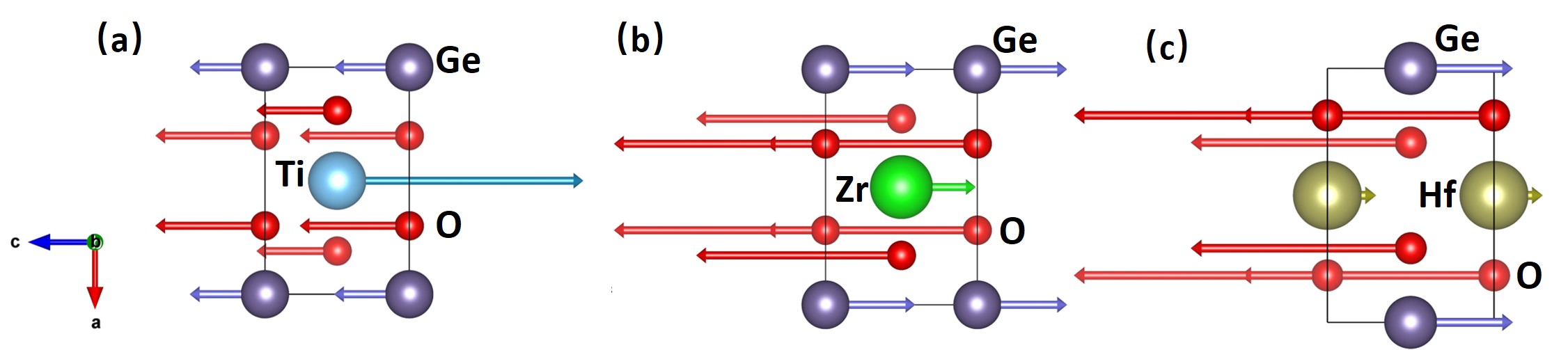}
\caption{\textbf{Representative vibrational modes and their contributions to the total dielectric constants in different \textit{Cmmm}-GeBO$_4$ materials.}
(a) $Cmmm$-GeTiO$_4$ (frequency of 55 cm$^{-1}$), (b) GeZrO$_4$ (frequency of 275 cm$^{-1}$) and (c) GeHfO$_4$ (frequency of 297 cm$^{-1}$), respectively.} 
\label{fig:figure3}
\end{figure*}

To illustrate the components of the dielectric tensor at an atomic level, we chose $Cmmm$-Ge$B$O$_4$ as an example to analyze the dielectric contribution of each vibration mode in Table \ref{table4}. $Cmmm$-Ge$B$O$_4$ has 6 atoms in the primitive cell, which results in 18 phonon branches, of which only 8 infrared (IR) active modes have non-zero mode-oscillator strength and contribute to the ionic dielectric constant. In GeTiO$_4$, the acoustic modes are dominated by the vibration of the heaviest atom, Ge with a mass ($M$) of 73. The large dielectric constant of GeTiO$_4$ is mainly due to the lowest-frequency IR mode at 55 cm$^{-1}$, which is caused by low-frequency vibration of Ti atoms along $c$ axis (as seen in Figure \ref{fig:figure3}a). Whereas, this mode is absent in GeZrO$_4$ and GeHfO$_4$, because the heavy Zr ($M$ = 91) and Hf ($M$ = 178) atoms contribute to the acoustic mode. The other modes for GeZrO$_4$ and GeHfO$_4$ have much higher frequencies, see Figure \ref{fig:figure3}b,c as examples, which only contribute slightly to dielectric constants. Combined with the large value of  $Z^*_{33}$ of Ti, the factors including light weight, large displacement, and large Born effective charges of the Ti atom, collectively distinguish GeTiO$_4$ from other materials in the prototype. Rignanese et al. \cite{rignanese2007first} also found that the variation in Ti oxides differs distinctively from Hf and Zr oxides and can be attributed to the difference in interatomic force constants. Similar information has been listed for all structural entries reported in this work. We believe such comprehensive analysis can yield a better understanding of the structure-dielectric property relation.

\begin{table}[ht]
\centering
\caption{The computed Born effective charges in $Cmmm$ Ge$B$O$_4$} \label{table3}
\begin{tabular}{c|ccc|ccc|ccc}
\hline
     & \multicolumn{3}{c|}{GeTiO$_4$} & \multicolumn{3}{c|}{GeZrO$_4$} & \multicolumn{3}{c}{GeHfO$_4$} \\
atom & 11       & 22       & 33      & 11       & 22       & 33      & 11       & 22       & 33      \\\hline
Ge   & 5.78     & 4.54     & 4.45    & 5.31     & 3.98     & 4.66    & 5.13     & 3.86     & 4.65    \\
$B$ (Ti, Zr, Hf)    & 3.85     & 5.85     & 7.77    & 4.58     & 5.70     & 6.21    & 4.54     & 5.55     & 5.91    \\
O1   & -1.35    & -3.91    & -2.26   & -1.37    & -3.53    & -2.31   & -1.35    & -3.39    & -2.29   \\
O2   & -3.46    & -1.29    & -3.84   & -3.58    & -1.32    & -3.13   & -3.48    & -1.32    & -2.99  \\\hline
\end{tabular}
\end{table}

\begin{figure*}[ht]
\centering
\includegraphics[width=0.9\linewidth]{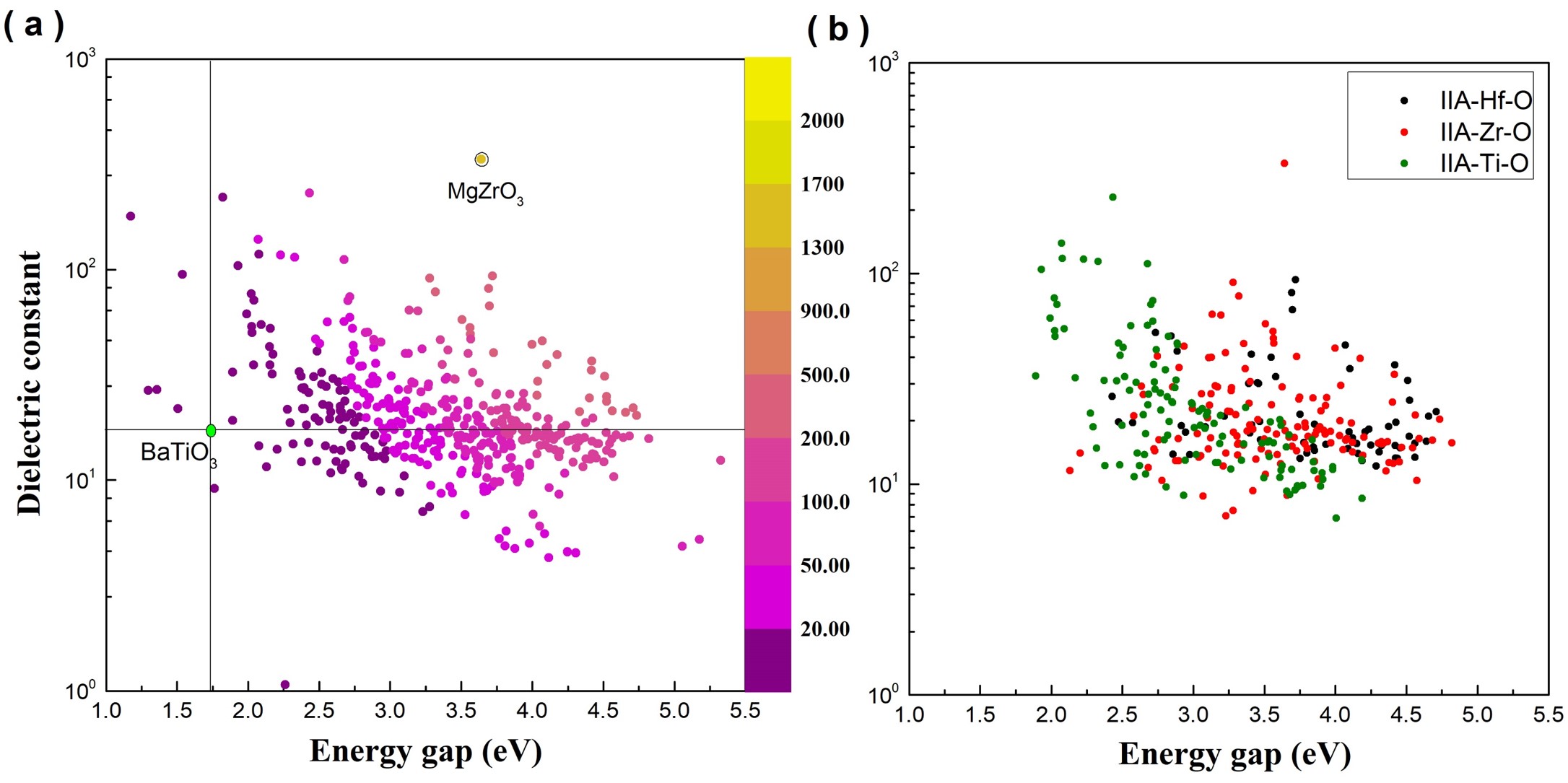}
\caption{\textbf{The correlation between dielectric constants and band gap in the calculated materials dataset.} (a) Dielectric constants and bandgap plots of A$_m$O$_n$-BO$_2$ (A: IIA/IIIA/IVA; B: IVB) oxides. Each point is color coded according to the fitness value; (b) Dielectric constants and bandgap plots of each individual system of IIA-IVB oxides.}
\label{fig:figure4}
\end{figure*}

\begin{figure*}[ht]
\centering
\includegraphics[width=0.65\linewidth]{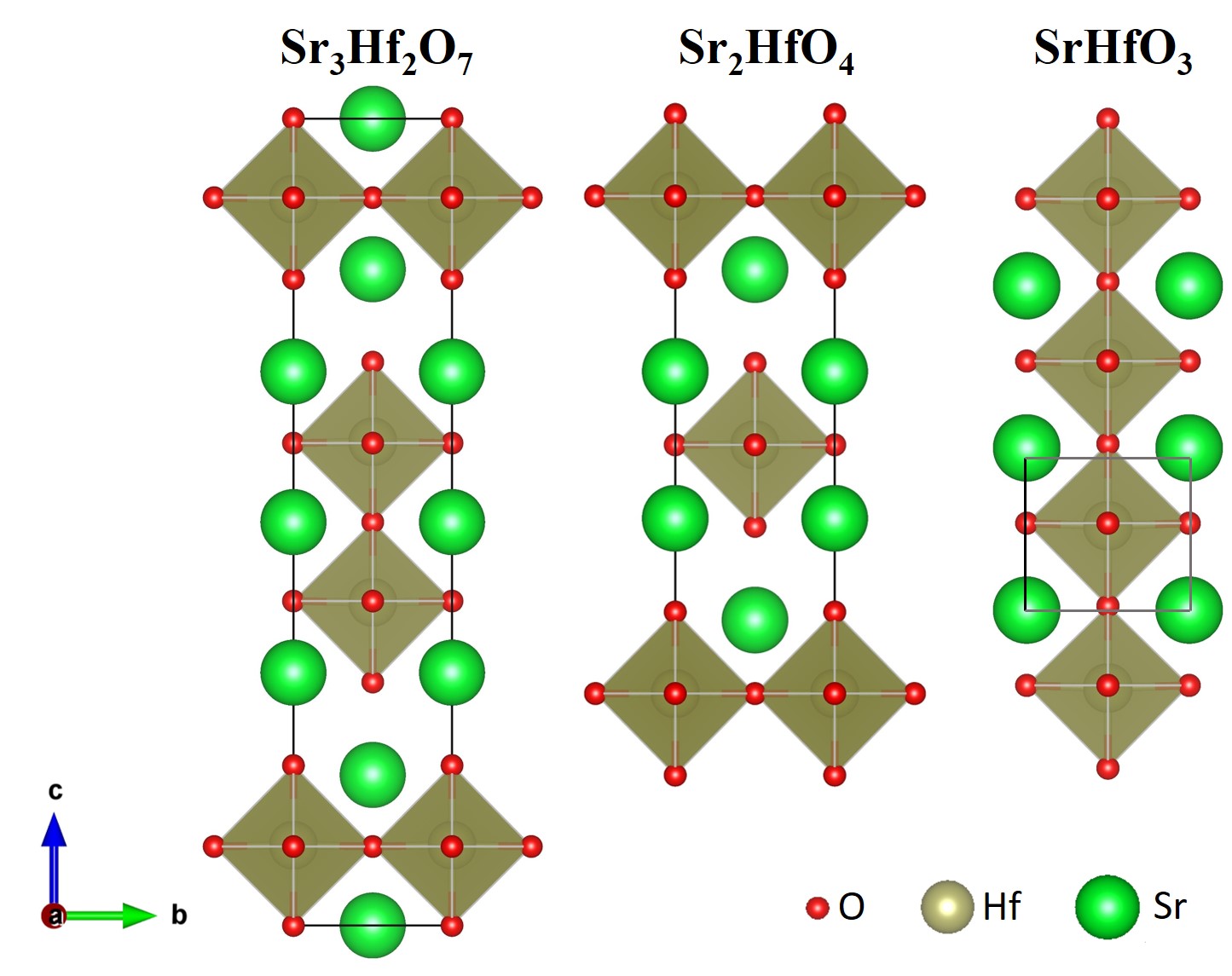}
\caption{\textbf{Representative high dielectric materials.} Crystal structures of $I4/mmm$-Sr$_3$Hf$_2$O$_7$ (ID = mp$-$779517), $I4/mmm$-Sr$_2$HfO$_4$ (ID = mp$-$13108) and $P4mm$-SrHfO$_3$ (ID = mp$-$768305), with dielectric constants of 135.00, 246.36 and 159.05, respectively, and bandgap of 3.68, 3.73 and 3.68 eV, respectively. }
\label{fig:figure5}
\end{figure*}

\begin{table}[ht]
\centering
	\caption{The computed ionic contribution to static dielectric constants decomposed to each IR active mode (cm$^{-1}$) in $Cmmm$ Ge$B$O$_4$.}
	\label{table-mode}
%\begin{tabular}{|c|c|c|c|c|c|c|c|c|c|c|c|}
\begin{tabular}{cccc|cccc|cccc}

\hline
\multicolumn{4}{c|}{GeTiO$_4$} & \multicolumn{4}{c|}{GeZrO$_4$}  & \multicolumn{4}{c}{GeHfO$_4$} \\ %\hline

Mode & $\epsilon_{11}$ & $\epsilon_{22}$ & $\epsilon_{33}$ & Mode & $\epsilon_{11}$ & $\epsilon_{22}$ & $\epsilon_{33}$ & Mode & $\epsilon_{11}$ & $\epsilon_{22}$ & $\epsilon_{33}$ \\ \hline
  573     & 2.1                      & 0                         & 0                         
& 554     & 2.4                      & 0                         & 0                         
& 556     & 2.2                      & 0                         & 0                   \\ %\hline
507       & 0                         & 3.0                      & 0                         
& 456     & 0                         & 3.7                      & 0                         
& 480     & 0                         & 3.3                      & 0                   \\ %\hline
493       & 0                         & 0                         & 0.6                      
& 419     & 0                         & 0                         & 0.3                      
& 446     & 0                         & 0                         & 0.3                \\ %\hline
361       & 0                         & 0.5                      & 0                        
& 320     & 4.5                      & 0                         & 0                         
& 316     & 7.1                      & 0                         & 0                   \\ %\hline
259       & 0                         & 0                         & 22.0                     
& 275     & 0                         & 0                         & 24.1                     
& 312     & 0                         & 1.2                      & 0                   \\ %\hline
238       & 22.6                     & 0                         & 0                         
& 268     & 7.9                      & 0                         & 0                         
& 297     & 0                         & 0                         & 18.1               \\ %\hline
222       & 0                         & 33.5                     & 0                         
& 202     & 0                         & 22.1                     & 0                         
& 246     & 3.9                      & 0                         & 0                   \\ %\hline
\textbf{55}  & 0                      & 0                         & \textbf{524.8}                    
& 149     & 0                         & 0                         & 0.1                      
& 200     & 0                         & 15.3                     & 0                   \\ \hline
 $\epsilon_0$     & 24.7             & 37.0                     & 547.5                   
&$\epsilon_0$     & 14.80             & 25.8                     & 24.5                     
&$\epsilon_0$     & 13.2             & 19.8                     & 18.4               \\ %\hline
 $\epsilon_\infty$    & 5.4          & 5.9                      & 7.0                  
&$\epsilon_\infty$    & 4.7          & 4.8                      & 5.3                     
&$\epsilon_\infty$    & 4.5          & 4.5                      & 5.0                \\ \hline
\end{tabular}
\label{table4}
\end{table}

\subsection*{High dielectrics}
%For optoelectronic and sensor applications, wide bandgap and high dielectric are required to the improvement of performance. 
The relation between the band gap and the dielectric constant of materials is depicted in Figure \ref{fig:figure4}. There exists a rough inverse correlation between these two properties, which is consistent with the previous reports \cite{yim2015novel, he2012review, naccarato2018designing}. Most of the oxides have bandgaps between 2 to 4.5 eV, while the dielectric constants range from 10 to 100. Hence most of them satisfy the requirements for gate electrode material, which usually require bandgap larger than 1 eV and dielectric constant larger than 3.9 (SiO$_2$) in a thin-film phase \cite{he2012review,kingon2000alternative}.

%\subsection*{Dielectric constant and bandgap of different element-based material}
The inverse relation of bandgap and dielectric is more obvious for each individual subset, especially in titania-based oxides. In Figure \ref{fig:figure4}b, 318 metastable or stable structures of IIAO-IVBO$_2$ were provided. The Ti oxides have lower bandgap than the others. Several Sr-Ti-O structures have dielectric constants larger than 100 while their bandgaps are lower than 3 eV, which obstruct the possible application of the device (CPU, DRAM or CMOS). No metastable or stable In-Hf-O structures were found. In IIIAO-IVBO$_2$ materials, $C$2/$m$-GaZr$_2$O$_5$ is the only material with both high $\epsilon$ (31.20) and $E_g$ (3.01 eV) values. Most of the ternary oxides have never been suggested to be promising high dielectrics, which opens an avenue for discovering novel high-k dielectrics. As for Si(Ge)O$_2$-Ti(Hf, Zr)O$_2$ ternaries, the inverse trend of E$_g$ and $\epsilon$ is apparent, three structures with $\epsilon >$ 90 were obtained, however, their E$_g$ are deficient. 

Our primary goal is to find a material with E$_g$ and $\epsilon$ larger than BaTiO$_3$, the leading material used in MLCC industry. BaTiO$_3$ transforms from the high-temperature paraelectric cubic phase ($Pm$-$3m$) to the low-temperature ferroelectric tetragonal phase ($P4mm$) at 406 K \cite{moriwake2010first}. The tetragonal polymorph has better ferroelectric, piezoelectric, and thermoelectric properties; hence, it is the most widely used polymorph in industry. Though the dielectric tensor is strongly dependent on temperature, we limited our study to materaisl at 0 K for simplicity. Most of the structures have both larger bandgaps and higher dielectric constants than tetragonal BaTiO$_3$ (E$_g$ = 1.72 eV, $\epsilon$ = 17.76). Among them, $I4/mmm$-Sr$_3$Hf$_2$O$_7$ achieves the highest fitness value ($\epsilon$ = 522.12 and E$_g$ = 3.68 eV). This hypothetical structure is also available in the open Materials Project (MP) database (mp$-$779517) without reported values on dielectric constants. However, after we checked the phonon, this structure is evidenced to be dynamically unstable due to the existence of imaginary phonon at X (1/2, 1/2, 0) and P (1/2, 1/2, 1/2) points. Similar conclusions were also mentioned in a recent study \cite{liu2019hybrid}. Further check the Sr-Hf-O system in MP database, two other structures, $P4mm$-SrHfO$_3$ and $I4/mmm$-Sr$_2$HfO$_4$ also stand out, with the dielectric constants of 246.36 and 159.05, respectively. Their crystal structures are very similar, as shown in Figure \ref{fig:figure5}. They both crystallize into tetragonal and have Hf centered octahedra. Their bonding length (2.06 \AA, 2.07 \AA\ and 2.06 \AA\ for Hf-O) and bandgap are almost the same. This result raises the possibility of looking for high dielectrics in these kinds of structures. In addition, $P2_1$/m-MgZrO$_3$ also shows high dielectric constant equals to 313.05, with bandgap of 3.64 eV. Given that DFT calculation systematically underestimates the fundamental bandgap \cite{perdew1983physical} by 10-40\% \cite{zeng2014evolutionary}, it is likely that $P2_1$/m-MgZrO$_3$'s band gap is larger than 4 eV, which can meet the requirement of application of CPU (4 eV) and other CMOS devices.

In the CPU and DRAM industry, dielectrics require $\epsilon >$ 30 for further device scaling \cite{yim2015novel}. With the additional limitation of E$_g >$ 3 eV for DRAM and E$_g >$ 4 eV for CPU we screened several materials listed in Table \ref{table5} (materials with calculated dielectric tensor in MP are not included). Most of the materials that meet the requirements are hafnia and zirconium based oxides. The structures are mainly monoclinic and orthorhombic. To evaluate high-performance dielectric materials for microelectronic device, we applied the fitness model according to Eq. \ref{fitness}. The fitness values are depicted by gradation of color in Figure \ref{fig:figure4}a. The bandgap can also significantly affect the fitness value, for example, $Pnma$-CaHfO$_3$ and $Pmc2_1$-Sr$_3$HfO$_5$ have similar dielectric constants (31.05 and 30.05). However, as a result of their bandgap difference (1.12 eV), their fitness values differ dramatically (323.21 for $Pnma$-CaHfO$_3$ and 91.06 for $Pmc2_1$-Sr$_3$HfO$_5$) 

\begin{table}%[ht]
\centering
	\caption{All materials satisfy $\epsilon > 30$, $E_g >$ 3 eV for DRAM and E$_g >$ 4 eV for CPU.}
	\begin{tabular}{cccccccc}
\hline
Formula	& Space group 	& E above hull 	& $\epsilon$ &	E$_g$ & fitness & MP-ID\\
\hline
CaHfO$_3$  & Pnma    & 0.02  & 31.05  & 4.51  & 323.21& mp-754853 \\
CaHfO$_3$  & P1    & 0.08  & 36.78  & 4.42  & 367.67& N/A\\
CaHfO$_3$  & Pmc2$_1$    & 0.07  & 35.36  & 4.10  & 304.59& N/A\\
CaHfO$_3$  & C2/c    & 0.04  & 45.77  & 4.07  & 387.86& N/A\\
SrHfO$_3$  & Imm2   & 0.02  & 93.26  & 3.72  & 494.40& N/A \\
SrHfO$_3$  & Pmc2$_1$   & 0.02  & 67.25  & 3.70  & 344.07& N/A \\
Sr$_2$HfO$_4$  & I4/mmm   & 0.03  & 81.17  & 3.69  & 411.92& N/A \\
Sr$_2$HfO$_4$  & Pbam  & 0.06  & 30.03  & 3.46  & 102.87& mp-752537\\
Sr$_3$HfO$_5$  & Pmm2    & 0.07  & 41.37  & 3.41  & 129.65& N/A \\
Sr$_3$HfO$_5$  & Pmc2$_1$   & 0.07  & 30.05  & 3.39  & 91.06& N/A \\
Ba$_3$Hf$_2$O$_7$  & I4/mmm   & 0.01  & 32.36  & 3.58  & 136.33& mp-754128\\
BaHfO$_3$  & Pm$\overline{3}$m    & 0.01  & 39.94  & 3.55  & 159.43& mp-998552\\
Mg$_2$ZrO$_4$  & Pc   & 0.07  & 33.18  & 4.42  & 331.05& N/A \\
MgZrO$_3$  & P2$_1$/m    & 0.06  & 313.05  & 3.64  & 1552.21& N/A  \\
BeZr$_6$O$_{13}$  & R3   & 0.02  & 39.44  & 4.17  & 351.72& N/A \\
BeZr$_4$O$_9$  & C2   & 0.03  & 52.98  & 3.56  & 216.58& N/A\\
CaZrO$_3$  & Pnma    & 0.00  & 44.18  & 4.00  & 361.19& mp-4571 \\
CaZr$_3$O$_7$  & Pmn2$_1$    & 0.00  & 57.74  & 3.51  & 214.64& N/A\\
Sr$_3$Zr$_2$O$_7$ &	I4/mmm	& 0.00	& 32.77	& 3.22&73.02 & mp-27690 \\
SrZrO$_3$  & C2/m   & 0.00  & 46.65  & 3.57  & 192.30& N/A\\
SrZrO$_3$  & Imma    & 0.00  & 49.28  & 3.56  & 202.18& mp-1080575\\
SrZrO$_3$  & Pnma    & 0.00  & 40.32  & 3.73  & 216.57& N/A\\
Sr$_4$ZrO$_6$  & C2   & 0.09  & 30.71  & 3.40  & 94.98& N/A \\
Sr$_3$Zr$_5$O$_{13}$  & Cm   & 0.04  & 35.31  & 3.37  & 102.72& N/A\\
SrZrO$_3$  & Pmc2$_1$    & 0.00  & 46.37  & 3.35  & 131.94& N/A\\
SrZrO$_3$  & Pmc2$_1$    & 0.00  & 78.20  & 3.32  & 209.15& N/A\\
SrZrO$_3$  & Cm    & 0.00  & 90.88  & 3.28  & 225.85& N/A \\
Sr$_2$ZrO$_4$  & Imm2    & 0.00  & 42.03  & 3.23  & 94.64& N/A\\
Sr$_3$ZrO$_5$  & Cm    & 0.10  & 63.51  & 3.20  & 135.10& N/A\\
SrZrO$_3$	& P4mm	   & 0.00  &  86.31 & 3.36 & 248.27&	mp-1068742 \\
BaZrO$_3$  & C2/c    & 0.00  & 64.04  & 3.13  & 121.14& N/A \\
Ba$_2$ZrO$_4$  & I4/mmm   & 0.00  & 39.86  & 3.11  & 71.80& mp-8335\\
Ga$_2$ZrO$_5$  & C2/m    & 0.07  & 31.20  & 3.01  & 46.13& N/A\\

\hline
\end{tabular}
\label{table5}
\end{table}

\section*{Technical Validation}

\begin{figure*}[ht]
\centering

\includegraphics[width=0.9\linewidth]{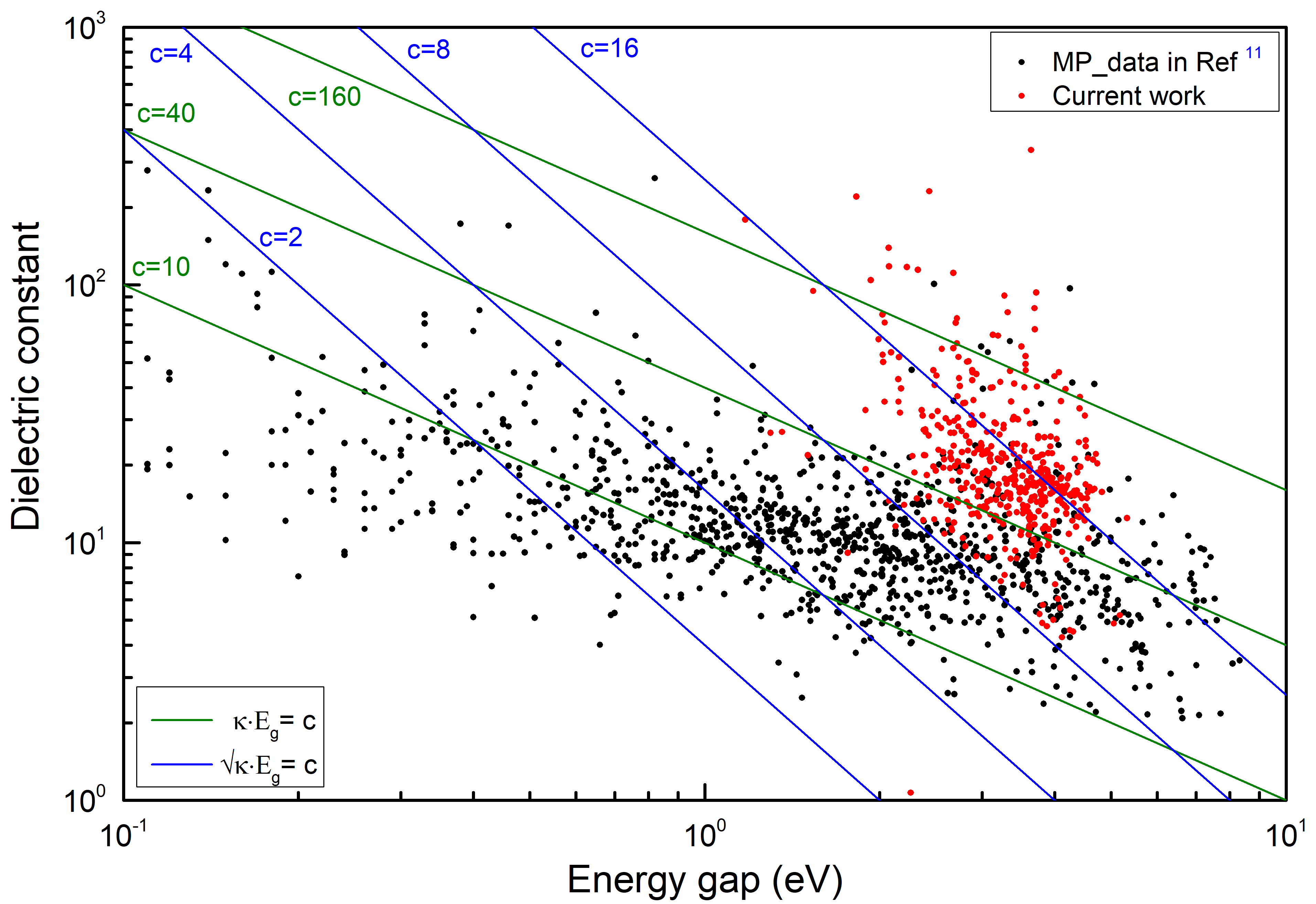}
\caption{\textbf{The distribution of high dielectric materials.} An updated map for the distribution of band gap and dielectric constants values results based on the results from this work (red) and Petousis \textit{et al} \cite{petousis2017high} (black).}
\label{fig:figure6}
\end{figure*}

In our calculation, 440 structures of ternary oxides were screened out, as shown in Figure \ref{fig:figure6}. Comparing with materials data in the Materials Project, our newly generated dataset has both larger bandgap and dielectric constant. Most of the ternary oxides are above $\epsilon*E_g = 40$ and $\surd \epsilon*E_g = 8$. We also selected a few representative materials to compare with Petousis's results \cite{petousis2017high}, for example, we conclude $\epsilon$\underline{\hbox to 2mm{}}poly and $\epsilon$ \underline{\hbox to 2mm{}}electronic of KSnSb are 15.60 and 12.54, which are in good agreement with the value of 15.55 and 12.47 in Ref. \cite{petousis2017high}. %Consistent results were also found for LaHBr$_2$, MnF$_2$ and Rb$_3$AuO. 
Since Petousis et al. have checked their computational error with experiments, and validated the dielectric constants of most materials deviate less than $\pm$25 \% from experiments, we believe our work should be as reliable as the previous work \cite{petousis2017high}. 

\section*{Usage Notes}

In this paper, we employed a first-principles crystal structure prediction method to perform a systematic structural study on a series of ternary oxides systems. For the low energy structures discovered from our prediction, we developed an automated computational screening scheme to evaluate their dielectric and electronic properties. This work generated a library of hypothetical materials which are promising for high dielectric applications. Among them, $P2_1$/m-MgZrO$_3$ achieves the best theorectical performance as it has both large bandgap (3.64 eV) and large dielectric constant (313.05). The rest 32 structures (such as $Pnma$-CaHfO$_3$, $Cm$-SrHfO$_3$ and $R3$-Zr$_6$BeO$_{13}$) with bandgap above 3 eV and dielectric constant above 30, may be useful in CPU or DRAM devices. Their structural, dielectric, and thermodynamic properties are archieved in a supplementary JSON file. In addition, we investigated the factors affecting the dielectric properties, pointing out that the dielectric properties are affected by multi-factors including vibration of atoms, Born effective charge, and atomic mass. Among these newly discovered structures, many of them were predicted from the first principles crystal structure prediction for the first time, suggesting that the crystal structure prediction methods as a complementary approach to the current high throughput screening based on data mining. The computational scheme developed here is entirely general to be used to search for other functional materials as well. 

\section*{Acknowledgements}

We acknowledge the support from China scholarship council (No.201706350087). The computing resources are provided by XSEDE (TG-DMR180040) and Center for Functional Nanomaterials under contract no. DE-AC02-98CH10086. Q.Z. and J.Q. thank Dr. V. Sharma for the useful discussions in dielectric calculation.

\section*{Author contributions}

Q.Z. designed the research. J.Q performed and analyzed the calculations. All authors contributed to interpretation and discussion of the data. Q.Z. and J.Q. wrote the manuscript.

\section*{Competing interests}

The authors declare no competing financial interests.

\bibliography{reference}

\end{document}